\documentclass[sigconf]{acmart}

\usepackage{tikz}

\usepackage{filecontents}
\usepackage{xspace}
\usepackage{tikz-cd}
\usepackage{pifont}
\usepackage{ctable} %
\usepackage{multirow}
\usepackage{listings}
\usepackage{stmaryrd}
\usepackage{comment}
\usepackage{semantic}
\usepackage{wrapfig}
\usepackage{multicol}
\usepackage{float} %

\newtheorem*{theorem*}{Theorem}

\usepackage{tikz}
\usetikzlibrary{calc}
\definecolor{proofob}{RGB}{32, 128, 32}
\definecolor{assump}{RGB}{32, 32, 224}
\definecolor{attackcol}{RGB}{192, 0, 0}
\definecolor{defendcol}{RGB}{0, 0, 192}
\newcommand{\trace}[2]{
\ifthenelse{\isempty{#1}}{\pi}{\pi_{#1}}
\ifthenelse{\isempty{#2}}{}{^{#2}}
}
\usepackage{xifthen}
\newcommand{\obseq}{\ensuremath{\approx_{\mathcal{L}}}}

\newcommand{\cerberus}{Cerberus\xspace}

\usepackage{url}
\usepackage{xurl} %

\usepackage[symbol]{footmisc}

\usepackage{afterpage}

\newcounter{daggerfootnote}

\newcommand{\etal}[0]{\textit{et al.}\xspace}
\newcommand{\sre}{SRE\xspace}
\newcommand{\tap}{TAP\xspace}
\newcommand{\tapc}{TAP$_C$\xspace}

\newcommand{\cloneop}{\texttt{Clone}\xspace}
\newcommand{\snapshotop}{\texttt{Snapshot}\xspace}
\newcommand{\launchop}{\texttt{Launch}\xspace}
\newcommand{\destroyop}{\texttt{Destroy}\xspace}
\newcommand{\enterop}{\texttt{Enter}\xspace}
\newcommand{\exitop}{\texttt{Exit}\xspace}
\newcommand{\pauseop}{\texttt{Pause}\xspace}
\newcommand{\resumeop}{\texttt{Resume}\xspace}

\newcommand{\sysclone}{\texttt{clone}\xspace}
\newcommand{\sysfork}{\texttt{fork}\xspace}
\newcommand{\uclid}{UCLID5\xspace}

\newcommand{\mdep}{\mathcal{M}^{EP}}
\newcommand{\mdam}{\mathcal{M}^{AM}_{PA}}
\newcommand{\mdap}{\mathcal{M}^{AM}_{perm}}
\newcommand{\mdev}{\mathcal{M}^{EV}}

\newcommand{\mdpc}{\mathcal{M}^{pc}}
\newcommand{\mdregs}{\mathcal{M}^{regs}}
\newcommand{\mdp}{\mathcal{M}^{paused}}
\newcommand{\mdiss}{\mathcal{M}^{IS}}
\newcommand{\mdcc}{\mathcal{M}^{CC}}
\newcommand{\mdrs}{\mathcal{M}^{RS}}
\newcommand{\mdpaf}{\mathcal{M}^{PAF}}

\newcommand{\tpc}{\mathit{pc}}
\newcommand{\tregs}{\Delta_{rf}}
\newcommand{\tmem}{\Pi}
\newcommand{\tam}{\mathit{a_{PA}}}
\newcommand{\tamp}{\mathit{a_{perm}}}
\newcommand{\tceid}{\mathit{e_{curr}}}
\newcommand{\towner}{\mathit{o}}
\newcommand{\temd}{\mathcal{M}}

\newcommand{\PA}{\mathit{PA}}
\newcommand{\VA}{\mathit{VA}}

\newcommand{\eqfontsize}{\small}
\usepackage{array,ragged2e}
\newcolumntype{P}[1]{>{\RaggedRight}p{#1}}

\newcommand\Intel{Intel\textsuperscript{\textregistered}\xspace}

\copyrightyear{2022}
\acmYear{2022}
\setcopyright{rightsretained}

\acmConference[CCS '22]{Proceedings of the 2022 ACM SIGSAC Conference on Computer and Communications Security}{November 7--11, 2022}{Los Angeles, CA, USA}
\acmBooktitle{Proceedings of the 2022 ACM SIGSAC Conference on Computer and Communications Security (CCS '22), November 7--11, 2022, Los Angeles, CA, USA}
\acmDOI{10.1145/3548606.3560595}
\acmISBN{978-1-4503-9450-5/22/11}

\usepackage{etoolbox}
\makeatletter
\patchcmd{\maketitle}{\@copyrightpermission}{
   \begin{minipage}{0.3\columnwidth}
     \href{https://creativecommons.org/licenses/by/4.0/}{\includegraphics[width=0.90\textwidth]{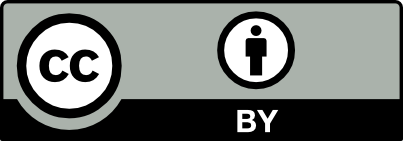}}
   \end{minipage}\hfill
   \begin{minipage}{0.7\columnwidth}
     \href{https://creativecommons.org/licenses/by/4.0/}{This work is licensed under a Creative Commons Attribution International 4.0 License.}
   \end{minipage}

   \vspace{5pt}
}{}{}

\makeatother

\begin{document}

\title[A Formal Approach to Enclave Memory Sharing]{Cerberus: A Formal Approach to \\ Secure and Efficient Enclave Memory Sharing}
\author{Dayeol Lee}
\authornote{Both authors contributed equally to the paper.}
\affiliation{\small{University of California, Berkeley}}
\email{dayeol@berkeley.edu}
\author{Kevin Cheang}
\authornotemark[1]
\affiliation{\small{University of California, Berkeley}}
\email{kcheang@berkeley.edu}
\author{Alexander Thomas}
\affiliation{\small{University of California, Berkeley}}
\email{alexthomas@berkeley.edu}
\author{Catherine Lu}
\affiliation{\small{University of California, Berkeley}}
\email{cathylu@berkeley.edu}
\author{Pranav Gaddamadugu}
\affiliation{\small{University of California, Berkeley}}
\email{pranavsaig@berkeley.edu}
\author{Anjo Vahldiek-Oberwagner}
\affiliation{\small{Intel Labs}}
\email{anjovahldiek@gmail.com}
\author{Mona Vij}
\affiliation{\small{Intel Labs}}
\email{mona.vij@intel.com}
\author{Dawn Song}
\affiliation{\small{University of California, Berkeley}}
\email{dawnsong@berkeley.edu}
\author{Sanjit A.~Seshia}
\affiliation{\small{University of California, Berkeley}}
\email{sseshia@berkeley.edu}
\author{Krste Asanović}
\affiliation{\small{University of California, Berkeley}}
\email{krste@berkeley.edu}

\renewcommand{\shortauthors}{Dayeol Lee et al.}

\ccsdesc[500]{Security and privacy~Formal security models}
\ccsdesc[500]{Security and privacy~Trust frameworks}
\ccsdesc[500]{Security and privacy~Trusted computing}

\keywords{enclaves, memory sharing, trusted execution environments; formal methods; formal verification; security; programming languages; computer architecture; Keystone; RISC-V; secure remote execution; }

\begin{abstract}

Hardware enclaves rely on a disjoint memory model, which maps each physical address to an enclave to achieve strong memory isolation.
However, this severely limits the performance and programmability of enclave programs.
While some prior work proposes enclave memory sharing, it does not provide a formal model or verification of their designs.
This paper presents \cerberus, a formal approach to secure and efficient enclave memory sharing.
To reduce the burden of formal verification, we compare different sharing models and choose a simple yet powerful sharing model.
Based on the sharing model, \cerberus extends an enclave platform such that enclave memory can be made immutable and shareable across multiple enclaves via additional operations.
We use \textit{incremental verification} starting with an existing formal model called the Trusted Abstract Platform (TAP).
Using our extended TAP model, we formally verify that \cerberus does not break or weaken the security guarantees of the enclaves despite allowing memory sharing. More specifically, we prove the Secure Remote Execution (\sre) property on our formal model.
Finally, the paper shows the feasibility of \cerberus by implementing it in an existing enclave platform, RISC-V Keystone.

\end{abstract}

\maketitle

\section{Introduction}

The hardware enclave~\cite{sgx,keystone,sanctum,komodo,trustzone,amd-sev,amd-sev-es} is a promising method of protecting a program~\cite{graphene-sgx, asylo, enclavedb, sgx-ml} by allocating a set of physical addresses accessible only from the program.
The key idea of hardware enclaves is to isolate a part of physical memory by using hardware mechanisms in addition to a typical memory management unit (MMU).
The isolation is based on a \textit{disjoint memory assumption}, which constrains each of the isolated physical memory regions to be owned by a specific enclave.
A hardware platform enforces the isolation by using additional in-memory metadata and hardware primitives.
For example, Intel SGX maintains per-physical-page metadata called the Enclave Page Cache Map (EPCM) entry, which contains the enclave ID of the owner~\cite{sgx-explained}.
The hardware looks up the entry for each memory access to ensure that the page is accessible only when the current enclave is the owner.

However, the disjoint memory assumption also significantly limits enclaves in terms of their performance and programmability.
First, the enclave needs to go through an expensive initialization whenever it launches because the enclave program cannot use shared libraries in the system nor clone from an existing process~\cite{pie}.
Each initialization consists of copying the enclave program into the enclave memory and performing \textit{measurements} to stamp the initial state of the program. 
The initialization latency proportionally increases depending on the size of the program and the initial data.
Second, the programmer needs to be aware of the non-traditional assumptions about memory. 
For instance, system calls like \sysfork or \sysclone no longer rely on efficient copy-on-write memory, resulting in significant performance degradation~\cite{graphene-sgx,asylo}.

A few studies have proposed platform extensions to allow
memory sharing of enclaves.
Yu~\etal~\cite{elasticlave} proposes Elasticlave, which modifies the platform such that each enclave can own multiple physical memory regions that the enclave can selectively share with other enclaves.
An enclave can map other enclaves' memory regions to its virtual address space by making a request, followed by the owner granting access.
Elasticlave improves the performance of enclave programs that relies on heavy inter-process communication (IPC).
Li~\etal~\cite{pie} proposes Plug-In Enclave (PIE), which is an extension of Intel SGX.
PIE enables faster enclave creation by introducing a \textit{shared enclave region}, which can be mapped to another enclave by a new SGX instruction \texttt{EMAP}.
\texttt{EMAP} maps the entire virtual address space of a pre-initialized \textit{plug-in} enclave.
PIE improves the performance of enclave programs with large initial code and read-only data (e.g., serverless workloads).
Although the prior work shows that memory sharing can substantially improve performance, they do not provide formal guarantees about security.

Unsurprisingly, the disjoint memory assumption of enclaves is crucial for the security of the enclave platforms.
Previous studies~\cite{tap, serval, komodo, tdxattestfv} formally prove high-level security guarantees of enclave platforms such as non-interference properties, integrity, and confidentiality based on the disjoint memory assumption.
However, to our best knowledge, no model formally verifies the security guarantees under the weakened assumption that the enclaves can share memory.
 
Practical formal verification requires choosing the right level of abstraction to model and apply automated reasoning.
Verification on models that conform to the low-level implementation~\cite{serval} or source-level code~\cite{sel4, cogent, cogent2, sewell2013translation, moat} is often platform-specific in that it only provides security guarantees to those implementations and thus does not apply generally.
If one seeks to verify that a memory-sharing approach on top of a family of enclave platforms is secure, it is not easy to reuse verification efforts for specific implementations.
We seek an approach that is incremental and also applicable to existing platforms.

Moreover, there are many ways one could design a memory-sharing model, each varying in complexity and flexibility.
Complex models can provide more flexibility to optimize the applications for performance, but this often comes at the cost of increasing the complexity of formal verification.
However, if memory sharing is too restrictive, it also becomes hard for programmers to leverage it for performance improvements.
Thus, we seek a simple sharing model with a balance between flexibility and ease of verification.

To this end, this paper presents \textit{\cerberus}, a formal approach to secure and efficient enclave memory sharing.
\cerberus chooses \textit{single-sharing} model with \textit{read-only} shared memory, which allows each enclave to access only one read-only shared memory.
We show that this design decision significantly reduces the cost of verification by simplifying invariants, yet still provides a big performance improvement for important use cases.
We formalize an enclave platform model that can accurately capture high-level semantics of the extension 
and formally verify a property called {\em Secure Remote Execution} (SRE)~\cite{tap}.
We perform \textit{incremental verification} by starting from an existing formal model called Trusted Abstract Platform (TAP)~\cite{tap} for which the SRE property is already established.
Finally, the paper shows the feasibility of \cerberus by implementing it in an existing platform, RISC-V Keystone~\cite{keystone}.
\cerberus can substantially reduce the initialization latency without incurring significant computational overhead.

To summarize, our contributions are as follows:
\begin{itemize}
    \item Provide a \textit{general} formal enclave platform model with memory sharing that weakens the disjoint memory assumption and captures a family of enclave platforms
    \item Formally verify that the modified enclave platform model satisfies \sre property via automated formal verification
    \item Provide programmable interface functions that can be used with existing system calls
    \item Implement the extension on an existing enclave platform and demonstrate that \cerberus reduces enclave creation latency
\end{itemize}
\section{Motivation and Background}
\label{sec:related_work} 

\subsection{Use Cases of Memory Sharing in Enclaves}
\label{sec:use_cases_of_sharing}
Many programs these days take advantage of sharing their memory with other programs.
For example, \textit{shared libraries} allow a program to initialize faster with less physical memory than static libraries because the operating system can reuse in-memory shared libraries for multiple processes.
Similarly, sharing large in-memory objects (e.g., an in-memory key-value store) can be shared across multiple processes.
Running a program inside an enclave disables memory sharing because of the disjoint memory assumption.
This section introduces a few potential use cases of memory sharing in enclaves to motivate \cerberus.
Memory sharing can significantly improve the performance of enclave programs that require multiple isolated execution contexts with shared initial code and data.

\paragraph{Serverless Workloads.} Serverless computing is a program execution model where the cloud provider allocates and manages resources for a function execution on demand.
In the model, the program developer only needs to write a function that runs on a language runtime, such as a specific version of Python.
Many serverless frameworks~\cite{fissionio,openfaas} reduce the cold-start latency of the execution with pre-initialized \textit{workers} containing the language runtime.
As described earlier by Li~\etal~\cite{pie}, the workers will suffer from an extremely long initialization latency (e.g., a few seconds) when they run in enclaves, as the language runtimes are typically a few megabytes (e.g., Python is 4~MB).
Because the difference between worker memories (e.g., heap, stack, and the function code) can be as small as a few kilobytes, a large amount of initialization latency and memory usage can be saved by sharing memory.

\paragraph{Inference APIs.} Machine learning model serving frameworks~\cite{huggingface,sagemaker,ray-serve} allow users to send their inputs and returns the model's inference results.
Serving different users with separate enclaves will have a longer latency as the model size increases.
As of now, the five most popular models in Huggingface~\cite{huggingface} have a number of parameters ranging from a few hundred million to a few hundred billion, which would occupy at least hundreds of megabytes of memory.
Sharing memory will drastically reduce the latency and memory usage of such inference APIs in enclaves.

\paragraph{Web Servers.} Multi-processing web servers handle requests with different execution contexts while sharing the same code and large objects.
For example, a web server or an API server that provides read access to a large object (e.g., front-end data or database) will suffer from long latency and memory usage running in an enclave.
If enclaves can share a memory, they can respond with lower latency and smaller memory usage.

\subsection{The Secure Remote Execution Property}
As mentioned earlier, much of the prior work identifies integrity and confidentiality as key security properties for enclave platforms. As a result, we aim to prove a property that is at least as strong as these two, which is the \sre property. To provide intuition behind the property, the typical setting for an enclave user is that the user wishes to execute their enclave program securely on a remote enclave platform. The remote platform is largely untrusted, with an operating system, a set of applications, and other enclaves that may potentially be malicious. Thus it is desirable to create a secure channel between the enclave program and the user in order to set up the enclave program securely. Consequently, in order to have end-to-end security, we need to ensure that the enclave platform behaves in the following three ways:
\begin{itemize}
    \item The measurement of an enclave on the remote platform can guarantee that the enclave is set up correctly and runs in a deterministic manner
    \item Each enclave program is integrity-protected from the untrusted entities and thus executes deterministically,
    \item Each enclave program is confidentiality-protected to avoid revealing secrets to untrusted entities.
\end{itemize}
These three behaviors manifest as the secure measurement, integrity, and confidentiality properties as defined in Section~\S\ref{sec:guarantees} and are ultimately what we guarantee for our platform model extended with \cerberus.

\subsection{Formal Models of Enclave Platforms}

Prior work has formally modeled and verified enclave platform models for both functional correctness and adherence to safety properties similar to the \sre property.
While verification at the source code level (e.g., Komodo~\cite{komodo}) provides proofs of functional correctness and noninterference of enclaves managed by a software \textit{security monitor}, existing verification efforts are often closely tied to the implementation, making it difficult to apply existing work to our extension.
A binary- or instruction-level verification (e.g., Serval~\cite{serval}) on the other hand, focuses on automating the verification of the implementation.
Working with binary-level models is often difficult and tedious because the binary often lacks high-level program context (e.g., variable names).
This paper aims to verify the enclave memory sharing on general enclave platforms. 
Thus, binary-level verification is a non-goal of this paper, while it can complement the approach by verifying that a given implementation refines our model at the binary level.

The {\em Trusted Abstract Platform model}~\cite{tap} is an abstraction of enclave platforms that was introduced with the \sre property. 
The \sre property states that an enclave execution on a remote platform follows its expected semantics and is confidentiality-protected from a class of adversaries defined along with the \tap model.
This property provides the end-to-end verification of integrity and confidentiality for enclaves running on a remote platform. 
It has also been formally proven that the state-of-the-art enclave platforms such as Intel's SGX~\cite{sgx,sgx-explained} and MIT's Sanctum~\cite{sanctum,sanctorum} refine the \tap model and hence satisfy \sre against various adversary models.
To our best knowledge, the \tap is the only model for formal verification that has been used to capture enclave platforms in a general way. The level of abstraction also makes it readily extensible. For these reasons, this paper extends the \tap model.
\section{Design Decisions} %

Several design decisions were made in our approach to conform to our design goals. The memory sharing model and interface designs are crucial for modeling, verification, and implementation. This section discusses the details of how we chose to design the memory-sharing model and interface.

\subsection{Writable Shared Memory}
\label{sec:writable_shared_memory}

Some programs use shared memory for efficient inter-process communication (IPC), which requires any writes to the shared memory to be visible to the other processes.
Elasticlave~\cite{elasticlave} allows an enclave to grant write permissions for a memory region to the other enclaves such that they can communicate without encrypting or copying the data.
However, the authors also show that such \textit{writable shared memory} requires the write permission to be dynamically changed to prevent interference between enclaves.
As formal reasoning on memory with dynamic permission will introduce a non-trivial amount of complexity, we leave this direction as future work.
Thus, \cerberus does not support use cases based on IPC or other mutable shared data.
Similarly, PIE~\cite{pie} also only enables read-only memory sharing among enclaves.

\subsection{Memory Sharing Models}
\label{sec:dec_mem_share_models}

Fig.~\ref{fig:sharing_designs} shows four different memory-sharing models with varying levels of flexibility. 
We discuss the implications for the implementation and the feasibility of formal verification for each model.
For this discussion, we use the number of \textit{access relations} between enclaves and memory regions as a metric for the complexity of both verification and implementation.

\definecolor{physgroupgray}{gray}{0.4}
\begin{figure}[t]
    \centering
	\includegraphics[width=0.85\columnwidth]{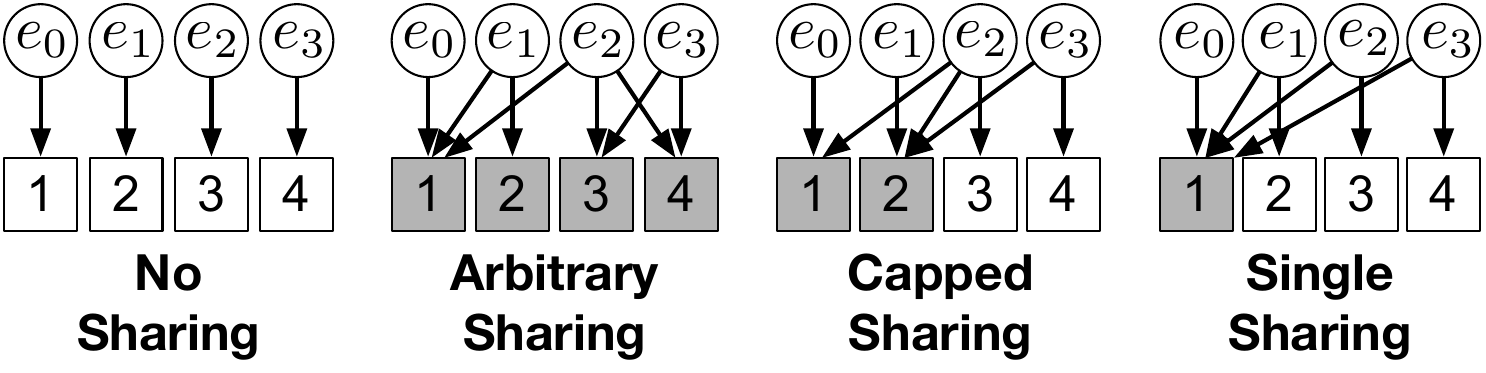}
	\caption{Memory sharing models with varying flexibility. \textcolor{physgroupgray}{Gray} (and white) boxes indicate shareable (and non-shareable) physical memory regions, and circles indicate enclaves. An edge from an enclave to physical memory is an \textit{access relation} stating that an enclave can access the memory it points to. The figure only depicts cases where the number of memory regions $m$ is the same as that of enclaves $n$, but $m$ can be greater than $n$ in practice.}
	\label{fig:sharing_designs}
\vspace{-1em}
\end{figure}

\paragraph{\textbf{No Sharing}}
We refer to the model that assumes the disjoint memory assumption as the \textit{no-sharing} model, which is implemented in state-of-the-art enclaves~\cite{sgx, keystone, komodo}.
The no-sharing model strictly disallows sharing memory and assigns each physical address to only one enclave.
As a result, the number of access relations is $O(max(m,n))=O(m)$, where $m$ is the number of physical memory regions, and $n$ is the number of enclaves. 
Thus, implementations with no-sharing will require metadata scaling with $O(m)$ to maintain the access relations.
For instance, each SGX EPC page has a corresponding entry in EPCM, which contains the owner ID of the page.
The no-sharing model has been formally verified at various levels~\cite{tap,komodo,serval}. 

\paragraph{\textbf{Arbitrary Sharing}}

One can completely relax the sharing model and allow any arbitrary number of enclaves to share memory (as in Elasticlave).
We refer to this sharing model as the \textit{arbitrary-sharing} model.
In this case, the number of access relations between enclaves is $O(mn)$. 
Consequently, arbitrary sharing requires metadata scaling with $O(mn)$.

\paragraph{\textbf{Capped Sharing}}

To achieve scalability in the number of access relations, one can constrain the sharing policy such that each enclave can only access a limited number of shared physical memory regions.
We refer to this sharing model as the \textit{capped-sharing} model.
In Fig.~\ref{fig:sharing_designs}, capped sharing shows an example where each enclave is only allowed to access at most two additional shared physical memory regions.
As an example, PIE~\cite{pie} introduces a new type of enclave called \textit{plug-in} enclave, which can be mapped to the virtual address space of a normal enclave.
This reduces the number of relations to $O(kn + m)$, where $k$ is the number of shared physical memory regions that are allowed to be accessed by an enclave.

\paragraph{\textbf{Single Sharing}}
\label{sec:single_sharing}
\begin{figure}
\includegraphics[width=0.8\columnwidth]{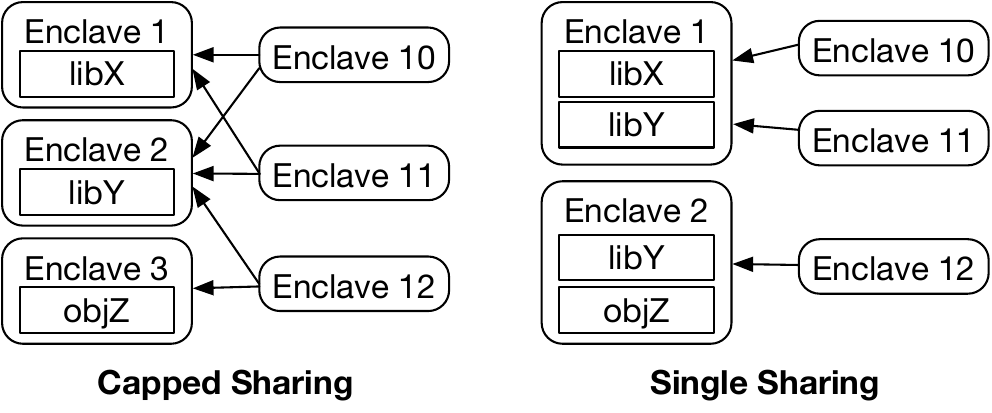}
\caption{Difference between capped- and single-sharing models in use cases. \texttt{libX}, \texttt{libY}, and \texttt{objZ} are the large libraries or objects that enclaves want to share. Enclave 10 and 11 relies on \texttt{libX} and \texttt{libY}, while Enclave 12 relies on \texttt{libY} and \texttt{objZ}.}
\label{fig:capped_vs_single}
\vspace{-0.5em}
\end{figure}

The \textit{single-sharing} model is a special case of the capped sharing with $k=1$. Thus, the model only allows enclaves to access the shared memory regions of a particular enclave.
Single sharing reduces the complexity to $O(m)$.

\subsection{Formal Verification of Sharing Models}

Formally verifying arbitrary- or capped-sharing models are challenging due to the flexibility of the models.
Verifying security properties such as \sre requires reasoning about safety properties with multiple traces and platform invariants with nested quantifiers.
In our experience, modeling an arbitrary number of shared memory would add to this complexity.
For example, one inductive invariant needed to prove \sre on \tap is that if a memory region is accessible by an enclave, the region is owned by the enclave.
To allow an arbitrary number of memory regions to be shared, the invariant should be extended such that it existentially quantifies over all relations, for example, stating that the owner of the memory is one of the enclaves that shared their memory with the enclave (See \S\ref{sec:guarantees} and Eq.~(\ref{eq:memory_ownership}) for details).
The encoding of the invariant in \tap uses first-order logic with the theory of arrays and, in general, is not decidable~\cite{fol_undec}. 
As a result, the introduction of this quantifier further complicates the invariant.
Despite the limiting constraint in \textit{capped sharing}, a formal model capturing any arbitrary limit $k$ would still require modeling an arbitrary number of the shared memory as in the \textit{arbitrary sharing} scheme and face the same complication.

In contrast, the single-sharing model significantly reduces the efforts of formal reasoning and implementation.
First, the formal reasoning no longer requires the complex invariant because the memory accessible to an enclave either belongs to the enclave itself or only another enclave that is sharing memory.
Second, the implementation becomes much simpler as it requires only one per-enclave metadata to store the reference to the shared memory.
The platform modification also becomes minimal as it only checks one more metadata per memory access.

Despite its simplicity, the single-sharing model can still improve the performance of programs by having all of the shared contents (e.g., shared library, initial code, and initial data) in a shared enclave.
Figure~\ref{fig:capped_vs_single} depicts the difference between capped- and single-sharing models.
With the capped-sharing model, each shareable content can be initialized with a separate enclave, allowing each enclave to map up to $k$ different enclave memory regions (i.e., \textit{Plug-In} enclaves in PIE).
Single-sharing model only allows each enclave to map exactly one other enclave, leaving potential duplication in memory when heterogeneous workloads have shared code (e.g., \texttt{libY}).
We claim that the benefit of the model's simplicity outweighs the limitation, as the single sharing does not have notable disadvantages over capped-sharing when there is no common memory among heterogeneous workloads.

\subsection{Interface}

Enclave programs need interface functions to share memory based on the sharing model.
Elasticlave and PIE introduce explicit operations to \textit{map} or \textit{unmap} the shareable physical memory region to the virtual address space of the enclave.
For example, Elasticlave requires an enclave program to explicitly call \texttt{map} operation to request access to the region, which will be approved by the owner via \texttt{share} operation.
Similarly, PIE allows an enclave to use \texttt{EMAP} and \texttt{EUNMAP} instructions to map and unmap an entire plug-in enclave memory to the virtual address of the enclave.

Elasticlave and PIE allow an enclave program to map shareable physical memory regions to its virtual address space.
However, there are a few downsides to the approaches.
First, the programmers must manually specify which part of the application should be made shareable.
In most cases, the programmers must completely rewrite a program such that the shareable part of the program is partitioned into a separate enclave memory.
Second, a dynamic map or unmap requires local attestation, which verifies that the newly-mapped memory is in an expected initial state.
Thus, the measurement property of a program relies on the measurement property of multiple physical memory regions.

\cerberus takes an approach similar to a traditional optimization technique, which clones an address space with copy-on-write, as in system calls like \sysclone and \sysfork.
This approach fits \cerberus use cases where the shareable regions include text segments, static data segments, and dynamic objects (e.g., a machine learning model).
In general, programmers expect such system calls to copy the entire virtual address space of a process -- no matter what it contains -- to a newly-created process.
A similar interface will allow the programmers to write enclave programs with the same expectation.
Also, such an interface will not require additional properties or assumptions on measurements of multiple enclaves.
Since the initial code of an enclave already contains when to share its entire address space, the initial measurement implicitly includes all memory contents to be shared.

To this end, \cerberus introduces two enclave operations, which are \snapshotop and \cloneop.
\snapshotop freezes the entire memory state of an enclave, and \cloneop creates a logical duplication of an enclave.
We make \snapshotop only callable from the enclave itself, allowing the enclave to decide when to share its memory.
The adversary can call \cloneop any time, which does not break the security because it can be viewed as a special way of launching an enclave (See \S\ref{sec:cloneop}).
When the adversary calls \cloneop on an existing enclave, a new enclave is created and resumes with a copy-on-write (CoW) memory of the snapshot.
Thus, any changes to each of the enclaves after the \cloneop are not visible to each other.
The following sections formally discuss the sharing model and the interface of \cerberus.

\section{Formal Model}
\label{sec:approach}

We first introduce a threat model
in Section~\S\ref{sec:threatmodel} that is consistent with these goals
and the current state-of-the-art enclave threat models. 
Then, we list and justify our assumptions in Section~\S\ref{sec:assumptions}, 
introduce our formal models of the platform and adversary based on these assumptions
in Section~\S\ref{sec:tapc_model} and then introduce the two new operations \snapshotop
and \cloneop of \cerberus in Section~\S\ref{sec:new_ops}.

In section~\S\ref{sec:guarantees}, we use these formal models to define the
\sre~\cite{tap} property, which is a critical security property used to prove 
that enclaves executing in the remote platform are running as expected and confidentially.
These properties are then formally verified using incremental verification on \tap.
In other words, our formal models extend the \tap model introduced by Subramanyan~\etal~\cite{tap}.
While \sre has been proven on the extended \tap model, \cerberus design weakens the disjoint memory assumption to allow memory sharing.
In addition, it is not immediately clear that the two additional operations clearly preserve \sre. 
Thus, we prove that \sre still holds under our extended model with the operations. 
For the rest of the literature, we refer to the original formal platform model defined by Subramanyan~\etal~\cite{tap} as \tap and our extended model as \tapc.

\subsection{Threat Model}
\label{sec:threatmodel}

\begin{figure}[t]
    \centering
	\includegraphics[width=.8\columnwidth]{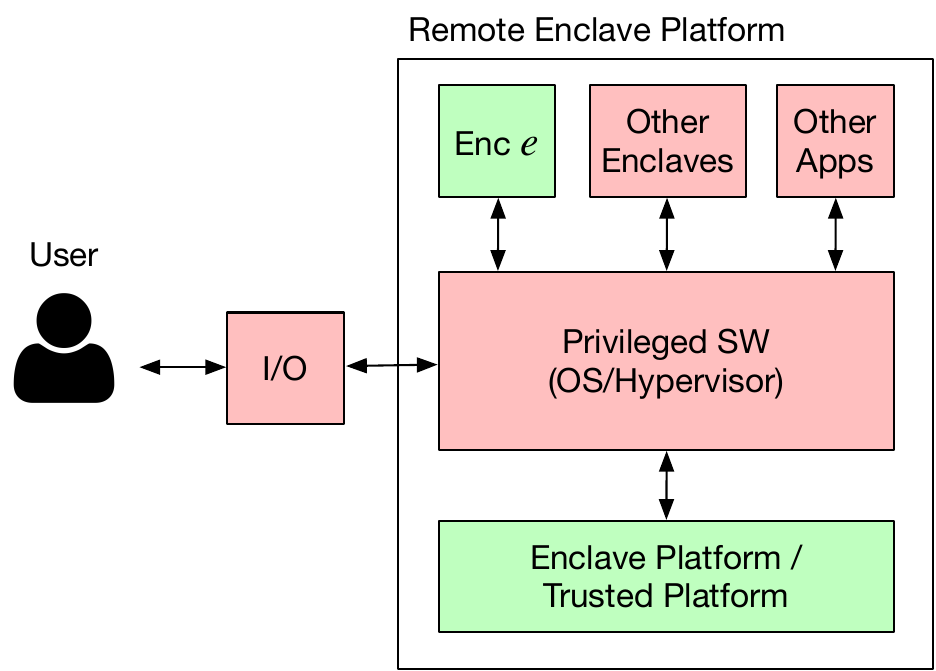}
	\caption{A user provisions their (protected) enclave $e$ in the remote enclave platform isolated from untrusted software. Green/red boxes indicate trusted/untrusted components.}
	\label{fig:user_example}
\vspace{-2em}
\end{figure}

Our extension follows the typical enclave threat model where the user's enclave program $e$ 
is integrity- and confidentiality-protected over the enclave states (e.g. register values and 
data memory owned by the enclave program) against any software adversary running in the remote enclave platform.
The software adversaries of an enclave include the untrusted operating system, 
user programs, and the other enclaves as shown in Figure~\ref{fig:user_example}.

With \cerberus, enclaves may share data or code that were 
common between enclaves before the introduction of the \cloneop.
We assume that the memory is implicitly not confidential among these enclaves with shared memory.
However, each enclave's memory should not be observable by the operating system or other enclaves and applications.
We ensure that the enclaves are still \textit{write-isolated}, which means that any modification to the data from one enclave
must not be observable to the other enclaves, even to the enclave that it cloned from.
Thus, any secret data needs to be provisioned after the enclave is cloned.
It is the enclave programmer's responsibility to make sure that the parent enclave does not contain any secret data that can be leaked through the children.

We do not consider the program running in the enclave to be 
vulnerable or malicious by itself.
For example, a program can generate a secret key in the shared memory,
and encrypt the confidential data of the child with the key. 
This would break confidentiality among children enclaves write-isolated from each 
other because the children will have access to the key in the shared memory.
We do not consider such cases, but this could be easily solved by having programs load secrets to their memory after they have created the distrusting children.

Since our main goal is to design a generic extension, we also do not 
consider any type of side-channel attack or architecture-specific attack~\cite{spectreishere, fallout, spectre, meltdown, foreshadow, lvi, ridl, controlled-channel,side-channel-cachezoom,side-channel-sgx-leaky,side-channel-conceal,membuster} in this paper.
We leave side-channel resilient interface design as future work.
We note that since the base \tap model has also been used to prove side-channel resiliency on some enclave platforms~\cite{keystone,sanctum}, it is not impossible to extend our proofs to such adversary models.
Denial of service against the enclave is also out of the scope of the paper; this is consistent with the threat models for existing state-of-the-art enclave platforms.

A formal model of the threat model is described in more detail in 
Section~\S\ref{sec:formal_adversary} after the formal definition of the platform.

\subsection{\tapc Model Assumptions}
\label{sec:assumptions}

Below, we summarize a list of assumptions about the execution model of \tapc
that we make for the purpose of simplifying and abstracting the modeling. These assumptions
are consistent with the adversary model described above:

\begin{itemize}
    \item \tapc inherits the assumptions and limitations of \tap~\cite{tap}, which include assuming that every platform and enclave operation is atomic relative to one another, assuming the DRAM is trusted, no support for demand paging, assuming a single-core and single-process model, and assuming properties of cryptographic functions used for measurement.
    \item If an enclave operation returns with an error code, we assume that the states of the platform are entirely reverted to the state prior to the execution of that operation. 
    \item State continuity of enclaves is out-of-scope in our models, consistent with prior work \tap~\cite{tap}, and can be addressed using alternative methods ~\cite{memoir,jangid}.
    \item The memory allocation algorithm (e.g. for copy-on-write) is deterministic given that the set of unallocated memory is the same. This means that given any two execution sequences of a platform, as long as the page table states are the same, the allocation algorithm will return the same free memory location to allocate.
\end{itemize}

Next, we introduce our formal models describing the platform 
which extends the existing \tap model with \snapshotop and \cloneop under these assumptions.

\subsection{Formal \tapc Platform Model Overview}
\label{sec:tapc_model}

As mentioned, a user of an \textit{enclave platform} typically has 
a program and data that they would like to run securely in a remote server, 
isolated from all other processes as shown in Figure~\ref{fig:user_example}. 
Such a program can be run as an enclave $e$. The remote server provides 
isolation using its hardware primitives and software for managing the 
enclaves, where the software component is typically firmware or a security monitor. 
This software component provides an interface for the enclave user through a 
set of operations, denoted by $\mathcal{O}$, for managing $e$. 
The goal is to guarantee that this enclave $e$ is protected from all other 
processes on the platform and running as expected. 
For the purpose of understanding the proofs, we refer to the enclave we 
would like to protect as the \textit{protected enclave} $e$. 
We make this distinction to differentiate it from adversary-controlled enclaves.

\subsubsection{\textbf{Platform and Enclave State}} The platform can be 
viewed as a transition system $M=\langle S, I, \rightsquigarrow \rangle$ that 
is always in some state denoted by $\sigma \in S$. Alternatively, $\sigma$ can be viewed as 
an assignment of values to a set of state variables $V$.
The platform starts in an initial state in the set $I$ and transitions
between states defined by a transition relation $\rightsquigarrow \subset S \times S$.
We write $(\sigma, \sigma^\prime) \in \rightsquigarrow$ to mean a valid transition of the
platform from $\sigma$ to $\sigma^\prime$. 
An execution of the platform 
therefore emits a (possibly infinite) sequence of states $\pi=\langle \sigma^0, \sigma^1, ...\rangle$,
where $(\sigma^i, \sigma^{i+1}) \in \rightsquigarrow$ for $i\in\mathbb{N}$. 
We write $\pi^i = \sigma^i$ interchangeably, but will usually write $\pi^i$ whenever 
referencing a specific trace. When an enclave is initially launched, 
it is in the initial state prior to enclave execution, which we indicate 
using the predicate $init_e(\sigma): S \rightarrow Bool$. 
We describe the set of variables $V$ and enclave state $E_e(\sigma)$ for \tapc
in the following Section~\S\ref{sec:tapc_state_vars}.

\subsubsection{\tapc State Variables}
\label{sec:tapc_state_vars}
Each of the variables $V$ in \tapc are shown in Table~\ref{fig:tap_c_state_vars}.
$\tpc: VA$\footnote{We write $v$: $T$ to mean variable $v\in V$ has type $T$} is an abstraction of the program counter whose value is a 
virtual address from the set of virtual addresses $VA$. $\tregs: \mathbb{N} \rightarrow W$ 
is a register file that is a map\footnote{of type $L \rightarrow R$, where 
the index type is $L$ and value type is $R$.} 
from the set of register indices (of natural numbers) $\mathbb{N}$ to the set of words $W$. 
$\tmem: PA \rightarrow W$ is an abstraction of memory that maps the set of 
physical addresses $PA$ to a set of words. We write $\tmem[a]$ to represent 
the memory value at a given physical address $a \in \PA$. A page table abstraction 
defines the mapping of virtual to physical addresses $\tam$ and access 
permissions $\tamp: VA \rightarrow ACL$, where $ACL$ is the set of read, write, and execute permissions. 
$ACL$ can be defined as the product $VA \rightarrow Bool \times Bool \times Bool$, 
where $Bool \overset{.}{=} \{ true, false \}$ and the value of the map corresponds to 
the read, write, and execute permissions for a given virtual address index\footnote{We use $\overset{.}{=}$ to mean \textit{by definition} to differentiate between the \textit{equality} symbol $=$.}. 
$\tceid: \mathcal{E}_{id}$ represents the current enclave that is executing. 
$\mathcal{E}_{id} = \mathbb{N} \cup \{ \mathcal{OS} \} \cup \{ e_{inv} \}$ is the set of enclave IDs 
represented by natural numbers and a special identifier $\mathcal{OS}$ representing 
the untrusted operating system. We reserve the identifier $e_{inv}$ to refer to the 
invalid enclave ID which can be thought of as a default value that does not refer 
to any valid enclave. For the ease of referring to whether an enclave is valid and launched, we define the predicate $valid(e_{id}) \overset{.}{=} e_{id} \neq e_{inv} \land e_{id} \neq \mathcal{OS}$ 
that returns whether or not an ID is a valid enclave ID. The \textit{active} predicate returns true for an enclave $e$ whenever it is launched or cloned and not yet destroyed in state $\sigma$.
$\towner$ is a map that describes the ownership of physical addresses, 
each of which can be owned by an enclave (with the corresponding enclave ID) or the untrusted operating system.

Lastly, each enclave $e$ has a set of enclave metadata $\temd$, 
which is a record of variables described in Table~\ref{fig:tap_c_metadata}. 
We abuse notation and write $\mdpc[e]$ to represent the program counter 
$\mdpc$ of $e$ in the record stored in the metadata map $\temd$. 
We use the enclave index operator $[\cdot]$ similarly for the other metadata fields defined in Table~\ref{fig:tap_c_metadata} to refer to a particular enclave's metadata. 
$\mdep[e]$ is the entry point of the enclave that the enclave $e$ starts in after the 
\launchop and before \enterop. 
$\mdam[e]$ is the virtual address map of the enclave program. 
$\mdap[e]$ is the map of address permissions for each virtual address. 
$\mdev[e]$ is the map from virtual addresses to Boolean values representing whether 
an address is allocated to the enclave. $\mdpc[e]$ is the current program counter of the enclave. 
$\mdregs[e]$ is the saved register file of the enclave. 
$\mdp[e]$ is a Boolean representing whether or not the enclave has been paused and is initially false at launch.

These variables were introduced in the base \tap model and are unmodified in \tapc. 
We introduce the remaining four metadata variables required for \cerberus in 
Section~\S\ref{sec:new_ops}, which are additional state variables in \tapc that are not defined in the base \tap model.

The state $E_e(\sigma)$ is a projection of the platform state to the 
enclave state of $e$ that includes $\mdep[e]$, $\mdam[e]$, $\mdap[e]$, $\mdev[e]$, $\mdpc[e]$, $\mdregs[e]$, and the 
projection of enclave memory $\lambda v\in VA. \textit{ITE}(\mdev[e][v], \tmem[\mdam[v]], \bot)$.  
In the last expression, $\lambda v\in \textit{VA}. E$ is the usual lambda operator 
over the set of virtual addresses $v$ and expression body $E$, $\textit{ITE}(c, expr_1, expr_2)$ 
is the if then else operator that returns $expr_1$ if condition $c$ is true and $expr_2$ otherwise.
$\bot$ is the constant bottom value which can be thought of as a don't-care or unobservable value. 
This projection of memory represents all memory accessible to enclave $e$, 
including shared memory and memory owned by enclave $e$ as 
referenced by the virtual address map $\mdam$. 

\begin{table}[tb]
\centering
    \footnotesize
    \centering
    \vspace{-0.25cm}
        \begin{tabular}[t]{ p{0.14\columnwidth} p{0.2\columnwidth}  p{0.46\columnwidth} }
        \specialrule{.1em}{.1em}{.1em}
        \textbf{State Var.} & \textbf{Type} & Description\\
        \hline
        $\tpc$ & $VA$ & The program counter.\\
        $\tregs$ & $\mathbb{N} \rightarrow W$ & General purpose registers.\\
        $\tmem$ & $PA \rightarrow W$ & Physical memory.\\
        $\tam$ & $VA \rightarrow PA$ & Page table abstraction; virtual to physical address map.\\
        $\tamp$ & $VA \rightarrow ACL$ & Page table abstraction; virtual to their permissions.\\
        $\tceid$ & $\mathcal{E}_{id}$ & Current executing enclave ID (or  $\tceid=\mathcal{OS}$ if the OS is executing). \\
        $\towner$ & $PA \rightarrow \mathcal{E}_{id}$ & Map from physical addresses to the enclave that owns it.\\
        $\temd$ & $\mathcal{E}_{id} \rightarrow \mathcal{E}_M$ & Map of enclave IDs to enclave metadata. $\texttt{emd}[\mathcal{OS}]$ stores a checkpoint of the OS.\\
        \specialrule{.1em}{.1em}{.1em}
        \end{tabular}
    \caption{\tapc State Variables $V$.}
    \label{fig:tap_c_state_vars}
    \vspace{-2em}
\end{table}
\begin{table}[tb]
    \footnotesize
    \centering
        \begin{tabular}[t]{ p{0.2\columnwidth} p{0.2\columnwidth} p{0.4\columnwidth} }
        \specialrule{.1em}{.1em}{.1em}
        \textbf{State Var.} & \textbf{Type} & \textbf{Description of each field}\\
        \hline
        $\mdep$ & $VA$ & Enclave entrypoint.\\
        $\mdam$ & $VA \rightarrow PA$ & Enclave's virtual address map.\\
        $\mdap$ & $VA \rightarrow ACL$ & Enclave's address permissions.\\
        $\mdev$ & $VA \rightarrow Bool$ & Set of private virtual addresses. \\
        $\mdpc$ & $VA$ & Saved program counter.\\
        $\mdregs$ & $\mathbb{N} \rightarrow W$ & Saved registers.\\
        $\mdp$ & $Bool$ & Whether enclave is paused.\\
        ${\mdiss}^\dagger$ & $Bool$ & Whether the enclave is a snapshot.\\
        ${\mdcc}^\dagger$ & $\mathbb{N}$ & Number of children enclaves.\\
        ${\mdrs}^\dagger$ & $\mathcal{E}_{id}$ & Enclave's root snapshot.\\
        ${\mdpaf}^\dagger$ & $PA \rightarrow Bool$ & Map of free physical addresses.\\
        \specialrule{.1em}{.1em}{.1em}
        \end{tabular}
    \caption{Records of \tapc $\mathcal{E}_M$ enclave metadata. $\dagger$ indicates additional state variables added to support \snapshotop \& \cloneop.}
    \label{fig:tap_c_metadata}
    \vspace{-2em}
\end{table}

\subsubsection{\textbf{Enclave Inputs and Outputs}}
Communication between an enclave $e$ and external processes for 
a given state $\sigma$ are controlled through $e$'s inputs $I_e(\sigma)$ and 
its outputs $O_e(\sigma)$. $I_e(\sigma)$ includes the arguments to the operations 
that manage enclave $e$, areas of memory outside of the enclave that the enclave 
may access and an untrusted attacker may write to, and randomness from the platform. 
$O_e(\sigma)$ contains the outputs of enclave $e$ that are writable to by $e$ and accessible to the attacker and the user.

\subsubsection{\textbf{Platform and Enclave Execution}}
An execution of an enclave $e$ is defined by the set of operations from $\mathcal{O}$, 
in which the execution of an operation is deterministic up to its input $I_e(\sigma)$ 
and current state $E_e(\sigma)$. This means that given the same inputs $I_e(\sigma)$ and 
enclave state $E_e(\sigma)$, the changes to enclave state $E_e(\sigma)$ is deterministic. 
The set of operations for the base \tap model is $\mathcal{O}_{base} \overset{.}{=} \{ \launchop, \destroyop, \enterop, \exitop, \pauseop, \resumeop, \texttt{AdversaryExecute} \}$. 
\tapc extends the base set with two additional operations: 
$\mathcal{O} \overset{.}{=} \mathcal{O}_{base} \cup \{ \snapshotop, \cloneop \}$. We use the predicate $curr(\sigma) = e$ to indicate that 
enclave $e$, which may be adversary controlled, is executing at state $\sigma$ and $curr(\sigma) = \mathcal{OS}$ to indicate 
that the operating system is executing.

\subsubsection{\textbf{Formal Adversary Model}}
\label{sec:formal_adversary}
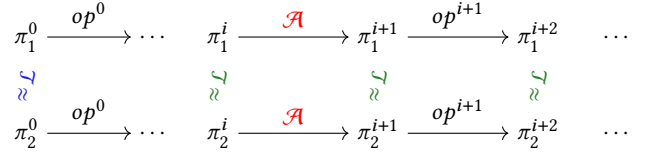
\begin{figure}[tbh]
    \begin{center}
        \tikzstyle{state}=[circle,draw=black]
        \newcommand{\tsblk}[1]{\normalsize{\ensuremath{#1}}}
        \newcommand{\tsr}[1]{\textcolor{proofob}{\normalsize{\ensuremath{#1}}}}
        \newcommand{\tsg}[1]{\textcolor{assump}{\normalsize{\ensuremath{#1}}}}
        \newcommand{\tsa}[1]{\textcolor{red}{\normalsize{\ensuremath{#1}}}}
        \newcommand{\tsd}[1]{\textcolor{black}{\normalsize{\ensuremath{#1}}}}
        \begin{tikzpicture}[xscale=1.7,yscale=0.8]
            \coordinate (yoff) at (0, 0.80);
            \coordinate (c0) at (0,0);
            \coordinate (c1) at (1.00, 0);
            \coordinate (c2) at (1.5, 0);
            \coordinate (c3) at (2.75, 0);
            \coordinate (c4) at (4.00, 0);
            \coordinate (c5) at (4.625, 0);
            \coordinate (c6) at (5.125, 0);
            \coordinate (c7) at (6.375, 0);
            \coordinate (c8) at (7.625, 0);

            \node (p0) at ($(c0) + (yoff)$) {\tsblk{\trace{1}{0}}};
            \node (q0) at ($(c0) - (yoff)$) {\tsblk{\trace{2}{0}}};
            \node[rotate=90] at (c0) {\tsg{\obseq}};
            \node (m0) at ($(c1) + (yoff)$) {\tsblk{\dots}};
            \node (n0) at ($(c1) - (yoff)$) {\tsblk{\dots}};
            \node[rotate=90] at (c2) {\tsr{\obseq}};
            \node (pi) at ($(c2)+(yoff)$) {\tsblk{\trace{1}{i}}};
            \node (qi) at ($(c2)-(yoff)$) {\tsblk{\trace{2}{i}}};
            \node (p1) at ($(c3) + (yoff)$) {\tsblk{\trace{1}{i+1}}};
            \node (q1) at ($(c3) - (yoff)$) {\tsblk{\trace{2}{i+1}}};
            \node[rotate=90] at ($(c3)$) {\tsr{\obseq}};
            \node (p2) at ($(c4) + (yoff)$) {\tsblk{\trace{1}{i+2}}};
            \node (q2) at ($(c4) - (yoff)$) {\tsblk{\trace{2}{i+2}}};
            \node[rotate=90] at (c4) {\tsr{\obseq}};
            \node (m1) at ($(c5) + (yoff)$) {\tsblk{\dots}};
            \node (n1) at ($(c5) - (yoff)$) {\tsblk{\dots}};

            \draw[->] (p0.east) -- node[above] {\tsd{op^0}} (m0.west);
            \draw[->] (q0.east) -- node[above] {\tsd{op^0}} (n0.west);
            \draw[->] (pi.east) -- node[above] {\tsa{\mathcal{A}}} (p1.west);
            \draw[->] (qi.east) -- node[above] {\tsa{\mathcal{A}}} (q1.west);
            \draw[->] (p1.east) -- node[above] {\tsd{op^{i+1}}} (p2.west);
            \draw[->] (q1.east) -- node[above] {\tsd{op^{i+1}}} (q2.west);
        \end{tikzpicture}
    \end{center}
    \caption{Illustrating the execution of two traces of the platform in the secure measurement, integrity and confidentiality proofs. Proof obligations for each property are checked as indicated by \textcolor{proofob}{\obseq} and equal initial condition indicated as \textcolor{assump}{\obseq}. $\textit{op}^i$ indicates enclave execution of an operation from $\mathcal{O}$ at step $i$ and $\textcolor{red}{\mathcal{A}}$ indicates an adversary execution.}
    \label{fig:traces_figure}
\end{figure}

In our model, untrusted entities such as the OS and untrusted enclaves are 
represented by an adversary $\mathcal{A}$ that can make arbitrary modifications 
to state outside of the protected enclave $e$, denoted by $A_e(\sigma)$. 
Consistent with the base \tap model, the untrusted entities and protected enclave $e$ 
takes turn to execute under interleaving semantics in our formal \tapc model, as 
illustrated in Figure~\ref{fig:traces_figure}. Under these semantics, the adversary
is allowed to take any arbitrary number of steps when an enclave is not executing.
Likewise, the protected enclave is allowed to take any number of steps when the adversary is
not executing without being observable to the adversary.

Conventionally, we define an adversary with an observation and tamper function 
that describes what the adversary can observe and change in the platform state 
during its execution to break integrity and confidentiality. The execution of the
adversary is the operation $\texttt{AdversaryExecute}$ in $\mathcal{O}_{base}$ during
which either the tamper or observation functions can be used by the adversary. 
Figure~\ref{fig:traces_figure} describes these two functions for our model.

\vspace{-5pt}

\paragraph{Tamper Function} The tamper function is used to model these 
malicious modifications to the platform state by the adversary and is defined over 
$A_e(\sigma)$ which includes any memory location that is not owned by the protected 
enclave $e$ and page table mappings. The semantics of the model allows the adversary 
to make these changes whenever it is executing. We allow all tampered states to be 
unconstrained in our models, which means they can take on any value. This type of
adversary tamper function over-approximates what the threat model can change
and is typically referred to as a havocing adversary~\cite{tap, vectre}.

\vspace{-4pt}

\paragraph{Observation Function} The adversary's observation function is denoted 
$obs_e(\sigma)$. In our model, we allow the adversary to observe locations of the memory 
that are not owned by the protected enclave $e$, described by the set 
$obs_e(\sigma) \overset{.}{=} O_e(\sigma) \overset{.}{=} \lambda p\in\PA.ITE(\sigma.\towner[p] \neq e, \sigma.\tmem[p], \bot)$. 
Intuitively, $obs_e$ is a projection of the platform state that is observable by 
the adversary whose differences should be excluded by the property. 
For example, if the same enclave program operating over different secrets 
reveals secrets through the output, that is a bug in the enclave program and we do not protect from this.
The adversaries may try to modify or read the enclave state during the lifetime of the enclave.

Under this threat model, we prove that the \tapc model still 
satisfies the \sre property described in Section~\S\ref{sec:guarantees}.

\subsection{The Extended Enclave Operations}
\label{sec:new_ops}

\cerberus is the extension of enclave platforms with two new operations 
\snapshotop and \cloneop to facilitate memory sharing among enclaves. 
Intuitively, \snapshotop converts the enclave executing the operation into
a read-only enclave and \cloneop creates a child enclave from the parent 
enclave being cloned so that the child enclave can read and execute the 
same memory contents as the parent at the time of clone.

This extension requires four new metadata state variables that are indicated 
in Table~\ref{fig:tap_c_metadata} with the $\dagger$ symbol. $\mdiss[e]$ is 
a Boolean valued variable indicating whether or not \snapshotop has been called 
on the enclave $e$. 
$\mdcc[e]$ is the number of children $e$ has, or in other words, 
the number of times a clone has been called on the enclave $e$ where $e$ 
is the parent of \cloneop. $\mdrs[e]$ is a reference to the root snapshot 
of $e$ if one exists, and $\mdpaf[e]$ is a map of addresses that have been 
assigned to $e$ but are not yet allocated memory.

We now define the semantics of the two new operations introduced in \cerberus.

\begin{figure}
	\centering
	\includegraphics[width=0.75\columnwidth]{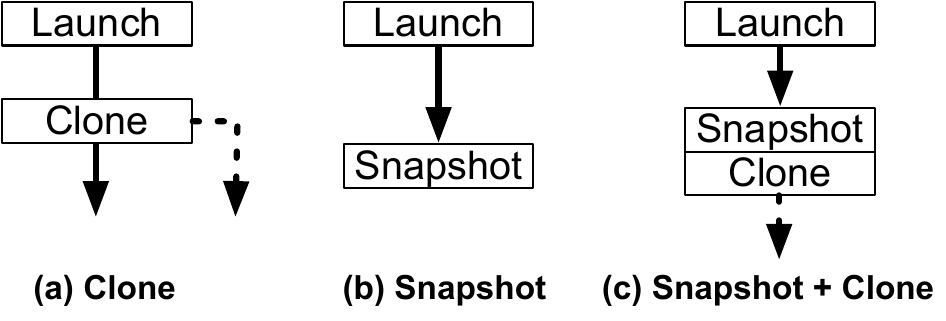}
	\caption{\cloneop, \snapshotop, and \cloneop-after-\snapshotop. The dotted arrow means an enclave newly-created by \cloneop.}
	\label{fig:clone_snapshot_basic}
\end{figure}

\subsubsection{\textbf{\cloneop}} 
\label{sec:cloneop}

\cloneop creates a clone of an existing logical 
enclave such that there exist two enclaves with identical enclave states. 
\cloneop alone provides a functionality similar to \sysfork and \sysclone system calls, no matter whether the platform enables memory sharing.
More concretely, the \cloneop takes in three arguments: the ID of the 
existing parent enclave $e^p_{id}\in\mathcal{E}_{id}$ to clone, the enclave ID of the child 
enclave $e^c_{id}\in\mathcal{E}_{id}$ and a set of physical addresses assigned to the 
child enclave $x_p \subset \PA$. 
The assigned physical addresses are marked as free (i.e., $\mdpaf[e_c][p] = true, \forall p \in x_p$), so that the parent's memory can be copied to them.
The child enclave $e_c$ with corresponding enclave ID $e^c_{id}$ is used to create a clone of the parent $e_p$ such that $E_{e_p}(\sigma) =  E_{e_c}(\sigma)$. In other words, the virtual memory of 
both enclaves are equal. 
We write $E_{e_p}(\sigma_0)$ to denote the initial state of the parent such that $init(E_{e_p}(\sigma))$.

We view the \cloneop as a special way of creating an enclave; 
instead of starting from the initial enclave state $E_{e_p}(\sigma_0)$, 
we start from an existing enclave $e_p$, which is effectively identical to 
creating two enclaves with the same initial state and then executing the 
the same sequence of inputs up until the point clone was called.

To prevent the malicious use of clones, we require the condition Eq.~\ref{eq:clone_success_cond} 
to hold during state $\sigma$ when \cloneop is called.
{\eqfontsize
    \begin{align}
        \label{eq:clone_success_cond}
        & \sigma.\tceid = \mathcal{OS} \land valid(e^p_{id}) \land active(e^p_{id}, \sigma) \quad \land\\\notag
        & valid(e^c_{id}) \land \neg active(e^c_{id}, \sigma) \quad \land\\\notag
        & e^c_{id} \neq e^p_{id} \quad \land\\\notag
        & \forall p\in\PA. p\in x_p \Rightarrow \sigma.\towner[p] = \mathcal{OS} \quad \land \\\notag
        & \textit{sufficient\_mem}(\sigma.\towner)
    \end{align} 
}
This condition states that the \cloneop succeeds if and only if the operating system (and hence not an enclave) is currently executing, the parent is a valid and active enclave, the child enclave ID is valid but it doesn't point to an active enclave, both the parent and child enclave IDs are distinct, all physical addresses in $x_p$ are owned by the OS (and thus can be allocated to the enclave), and there is \textit{sufficient memory} to be allocated to the enclave.

If the condition passes, \cloneop copies all of the data in the virtual address space of $e_p$ to $e_c$ to ensure write isolation.
For each virtual address $v$ mapped by $e_p$ (mapped), \cloneop first selects a physical address $p$ owned by $e_c$, copies the contents from $\tmem[\mdam[e_p][v]]$ to $\tmem[p]$, and update the page table of $e_c$ such that $\mdam[e_c][v] = p$.
This can be implemented in the platform itself (i.e., the security monitor firmware in Keystone) or in a local vendor-provided enclave (i.e., similar to the Quoting Enclave in \Intel SGX).

In Eq.~\ref{eq:clone_success_cond}, $\textit{sufficient\_mem}: \PA \rightarrow 
\mathcal{E}_{id} \rightarrow Bool$, \textit{sufficient\_mem} can be viewed as a predicate that determines whether there is enough memory to copy all data.
\textit{sufficient\_mem} is modeled abstractly in the \tapc model to avoid an expensive computation to figure out whether there is enough memory.

\cloneop is only called from the untrusted OS because it requires the OS to allocate resources for the new enclave.
Thus, if an enclave program needs to clone itself, it needs to collaborate with the OS to have it call \cloneop on its behalf.
As the newly-created enclave is still an isolated enclave, the SRE property on both parent and child enclaves should hold even with a malicious OS.  

\subsubsection{\textbf{\snapshotop}} 
\label{sec:snapshot}
\cloneop by itself still requires 
copying the entire virtual memory to ensure isolation.
To enable memory sharing, \snapshotop makes the caller enclave $e$ to be an immutable image (Figure~\ref{fig:clone_snapshot_basic}b).
After calling \snapshotop, $e$ becomes a special type of enclave referred to as a \textit{snapshot enclave} or the \textit{root snapshot} of its descendants. 
$e$ is no longer allowed to execute at this point because all of its memory becomes read- or execute-only. 
On the other hand, $e$ can be \textit{cloned} by \cloneop, where the descendants of $e$ are allowed to read directly from the $e$'s shared data pages.
Any writes from the descendants to physical addresses $p\in\PA$ owned by $e$ (i.e., $\sigma.\towner[p] = e$) trigger copy-on-write (CoW).
This scheme ensures that the descendant enclaves are still write-isolated from each other.

Like \cloneop, \snapshotop has a success condition described in 
Eq.~(\ref{eq:snapshot_success_cond}). The condition checks that 
the current executing enclave is valid $valid(\sigma.\tceid)$ and active $active(\sigma.\tceid, \sigma)$, 
$e$ is not already a snapshot and the enclave cannot have a root 
snapshot in the current state $\sigma$.
{\eqfontsize
    \begin{align}
        \label{eq:snapshot_success_cond}
        & \textit{valid}(\sigma.\tceid) \land \textit{active}(\sigma.\tceid, \sigma) \quad \land\\\notag
        & \neg\sigma.\mdiss[e] \land \neg\textit{valid}(\sigma.\mdrs[\sigma.\tceid])
    \end{align}
}
If \snapshotop is called successfully in a state that satisfies this condition, 
$e$ is marked as a snapshot enclave. In the formal model, the metadata 
state $\mdiss[e]$ is to \textit{true}.

\subsubsection{\textbf{\cloneop after \snapshotop}}

In order to make \cloneop work with \snapshotop,
\cloneop additionally increments $e_p$'s child count $\mdcc[e_p]$ by 1, and sets the root snapshot of $e_c$ (i.e., $\mdrs[e_c]$) to either $e_p$ or $e_p$'s root snapshot $\mdrs[e_p]$ if it has one.

\begin{figure}
    \centering
    \includegraphics[width=0.6\columnwidth]{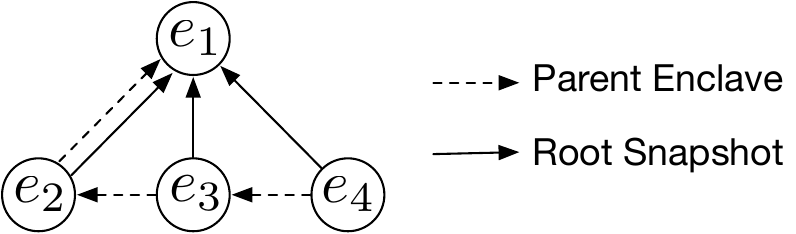}
    \caption{Parent-child relationship and root snapshot-child relationship of 
    four enclaves in Cerberus. Enclave $e_1$ is a snapshot and the parent enclave 
    of $e_2$, which is the parent of $e_3$, which is the parent of $e_4$. 
    Despite the nested parent relationship, the root snapshot of $e_2, e_3$, and $e_4$ are $e_1$.}
    \label{fig:max_1_tree_cerberus}
\end{figure}

With the single-sharing model, arbitrarily nested calls of \cloneop should still keep only one shareable enclave.
As shown in Fig.~\ref{fig:max_1_tree_cerberus}, there will be only one root snapshot $e_1$, whose memory is shared across all the descendants.
This means that even though cloning can be arbitrarily nested, the maximum height of the tree representing the root snapshot to the child enclave is one.

To maintain the same functionality, the virtual address space of the parent and the child should be the same right after \cloneop.
Thus, a descendant enclave memory will diverge from the shared memory when the descendant writes.
Unfortunately, there is no better way than having \cloneop copy the diverged memory from the parent to the child.
This is a limitation of \cerberus because the benefit of sharing memory will gradually vanish as the memory of the descendant diverges from the snapshot.
However, we claim that \cerberus is very effective when the enclaves mostly write to a small part of the memory while sharing the rest.
It is the programmer's responsibility to optimize their program by choosing the right place to call \snapshotop.

\section{Formal Guarantees}
\label{sec:guarantees}
To recap, one of our goals is to prove that our extension applied to an enclave platform does not weaken the high-level security property \sre. We accomplish this by reproving \sre on \tapc. As per Theorem 3.2~\cite{tap} (restated below as Theorem~\ref{thm:decomp}), it suffices to show that the triad of properties -- secure measurement, integrity, and confidentiality -- hold on \tapc to prove \sre. In this section, we formally define \sre, the decomposition theorem, and each of the properties along with informal justification as to why they hold in the \tapc model against the adversary described in Section~\S\ref{sec:approach}. Each of these properties has been mechanically proven on the base \tap model~\cite{tap} without \snapshotop and \cloneop, and in our work, we extend these proofs to provide the same guarantees for the memory adversary on the extended \tapc model. For brevity, we leave out some of the model implementation details and refer the reader to the GitHub repository for the proofs. We also provide a list of additional inductive invariants required to prove the properties in the \tapc model at the end of this section.

While there are several flavors of non-interference properties, the following properties are based on the observational determinism (OD)~\cite{hyperproperties, zdancewic03} definition of non-interference generalized for traces of concurrent systems. OD commonly shows up in several formalisms of confidentiality and integrity including the classic work on separability by Rushby~\cite{rushby} among other work~\cite{tap, SEC-008, vectre, HWSWcontracts}.
At a high level, OD states that if the initial states are low-equivalent and low inputs are the same, all states including intermediate states must also be low-equivalent (i.e., observationally deterministic functions of the low state/inputs). 
Whereas the classic non-interference property has an obligation~\cite{zdancewic03} to prove termination and does not reason about intermediate states. 
We find OD to be more appropriate because we desire to show that every state of execution is observationally deterministic and indistinguishable.

\subsection{Secure Remote Execution}
\begin{definition}[Secure Remote Execution]
Let $\pi=\{\sigma_0, \sigma_1, ...\}$ be a possibly unbounded-length sequence of platform states and $\pi^\prime=\{\sigma^\prime_0, \sigma^\prime_1, ...\}$ be the subsequence of $\pi$ containing all of the enclave executing states (i.e. $\forall i\in\mathbb{N}.curr(\sigma^\prime_i)=e$). Then the set $[\![ e ]\!] = \{ \langle I_e(\sigma^\prime_0), E_e(\sigma^\prime_0), O_e(\sigma^\prime_0) \rangle, ... | \textit{init}_e(E_e(\sigma_0))\}$ describes all valid enclave execution traces and represents the expected semantics of enclave $e$. A remote platform performs \sre of an enclave program $e$ if any execution trace of $e$ on the platform is contained within $[\![ e ]\!]$. In addition, the platform must guarantee that a privileged software attacker can only observe a projection of the execution trace defined by \textit{obs}.
\end{definition}

To prove \sre, the following theorem from prior work~\cite{tap} allows us to decompose the proof as follows.

\begin{theorem}\label{thm:decomp}
An enclave platform that satisfies secure measurement, integrity, and confidentiality property for any enclave program also satisfies secure remote execution.
\end{theorem}

\paragraph{\textbf{Secure Measurement}} In any enclave platform, the user desires to know that the enclave program running remotely is in fact the program that it intends to run. In other words, the platform must be able to \textit{measure} the enclave program to allow the user to detect any changes to the program prior to execution.  
The first part of the measurement property stated as Eq.~(\ref{eq:measure_1}) requires that the measurements $\mu(e_1)$ and $\mu(e_2)$ of any two enclaves $e_1$ and $e_2$ in their initial states are the same if and only if the enclaves have identical initial enclave states.
$\mu$ is defined to be the measurement function that the user would use to check that their enclave $e$ is untampered with in the remote platform.

\vspace{-0.3cm}
{\eqfontsize
    \begin{flalign}
    \label{eq:measure_1}
        \forall \sigma_1, \sigma_2 \in S. & \big( \mathit{init}(E_{e_1}(\sigma_1)) \land \mathit{init}(E_{e_2}(\sigma_2)) \big) \Rightarrow\\\notag
         &\big( \mu(e_1) = \mu(e_2) \iff E_{e_1}(\sigma_1) = E_{e_2}(\sigma_2) \big) &\\\notag
    \end{flalign}
}
\vspace{-0.7cm}

The second part of measurement ensures that the enclave executes deterministically given an initial state. This is formalized as Eq.~(\ref{eq:measure_2}), which states that any two enclaves $e_1$ and $e_2$ starting with the same initial states, executing in lockstep and with the same inputs at each step, should have equal enclave states and outputs throughout the execution. Together, these properties help guarantee to the user that their enclave is untampered with.

\vspace{-0.3cm}
{\eqfontsize
    \begin{align}
    \label{eq:measure_2}
        \forall \pi_1,\pi_2. 
        & \Big(E_{e_1}(\pi_1^0) = E_{e_2}(\pi_2^0) \quad \land\\\notag
        & \forall i\in\mathbb{N}.(\mathit{curr}(\pi_1^i)=e_1) \iff (\mathit{curr}(\pi_2^i)=e_2) \quad \land \\\notag
        & \forall i\in\mathbb{N}.(\mathit{curr}(\pi_1^i)=e_1) \Rightarrow  I_{e_1}(\pi_1^i) = I_{e_2}(\pi_2^i)\Big) \quad \Rightarrow \\\notag
        &\Big(\forall i\in\mathbb{N}. E_{e_1}(\pi_1^i) = E_{e_2}(\pi_2^i) \land O_{e_1}(\pi_1^i) = O_{e_2}(\pi_2^i)\Big) & \\\notag
    \end{align}
}
\vspace{-0.7cm}

With the addition of the \cloneop and \snapshotop, the measurement of enclaves does not change for two reasons. Eq.~(\ref{eq:measure_1}) is satisfied because the measurement of a child is copied over from the parent, and has an equivalent state as the parent. In addition, because each enclave child executes in a way that is identical to the parent without \cloneop, the child enclave $e_c$ is still deterministic up to the inputs $I_{e_c}(\sigma)$.

\paragraph{\textbf{Integrity}} The second property, integrity,  states that the enclave program's execution cannot be affected by the adversary beyond the use of inputs $I_e$ at each step and initial state $E_e(\pi_1^0)$, formalized as Eq.~(\ref{eq:integrity}).

\vspace{-0.3cm}
{\eqfontsize
    \begin{align}
    \label{eq:integrity}
        \forall \pi_1,\pi_2.
        & \Big(E_{e}(\pi_1^0) = E_{e}(\pi_2^0) \quad \land\\\notag
        & \forall i\in\mathbb{N}.(\mathit{curr}(\pi_1^i)=e) \iff (\mathit{curr}(\pi_2^i)=e) \quad \land \\\notag
        & \forall i\in\mathbb{N}.(\mathit{curr}(\pi_1^i)=e) \Rightarrow  I_{e}(\pi_1^i) = I_{e}(\pi_2^i)\Big) \quad \Rightarrow \\\notag
        &\Big(\forall i\in\mathbb{N}. E_{e}(\pi_1^i) = E_{e}(\pi_2^i) \land O_{e}(\pi_1^i) = O_{e}(\pi_2^i)\Big) \\\notag
    \end{align}
}
\vspace{-0.7cm}

\cloneop creates a logical copy of the enclave whose behavior matches the parent enclave had it not been cloned and thus clone does not affect the integrity of the enclave. \snapshotop freezes the enclave state and thus does not affect the integrity vacuously because the state of $e$ after calling snapshot does not change until its destruction.

\paragraph{\textbf{Confidentiality}} Lastly, the confidentiality property states that given the same enclave program with different secrets represented by $e_1$ and $e_2$ in traces $\pi_1$ and $\pi_2$ respectively, if the adversary starts in the initial state $A_{e_1}(\pi_1[0])$ and the protected enclave(s)  $e_1$ (and $e_2$) is operated with a (potentially malicious) sequence of inputs $I_{e_1}$, the adversary should not learn more than what's provided by the observation function $obs$ and hence its state $A_{e_1}(\sigma)$ and $A_{e_1}(\sigma)$ should be the same. The fourth line of Eq.~(\ref{eq:confidentiality}) requires that any changes by the protected enclave $e$ do not affect the observations made by the adversary in the next step. This is to avoid spurious counter-examples where secrets leak through obvious channels such as the enclave output which is a bug in the enclave program as explained in Section~\ref{sec:formal_adversary}.

\vspace{-0.3cm}
{\eqfontsize
    \begin{align}
    \label{eq:confidentiality}
        \forall \pi_1,\pi_2. 
        & \Big(A_{e_1}(\pi_1^0) = A_{e_2}(\pi_2^0) \quad \land\\\notag
        & \forall i\in\mathbb{N}.(\mathit{curr}(\pi_1^i)=\mathit{curr}(\pi_2^i) \land I_{e_1}(\sigma_1^i) = I_{e_2}(\sigma_2^i)) \quad \land \\\notag
        & \forall i\in\mathbb{N}.(\mathit{curr}(\pi_1^i)=e) \Rightarrow  \mathit{obs}(\pi_1^{i+1}) = \mathit{obs}(\pi_2^{i+1})\Big) \quad \Rightarrow \\\notag
        &\Big(\forall i\in\mathbb{N}. A_{e_1}(\pi_1^i) = A_{e_2}(\pi_2^i)\Big) & \\\notag
    \end{align}
}
\vspace{-0.7cm}

\snapshotop alone clearly does not affect the confidentiality of the enclave. \cloneop on the other hand also does not affect confidentiality because it creates a logical duplicate of an enclave. Had the adversary been able to break the confidentiality of the child $e_c$, it should have been able to break the confidentiality of the parent $e_p$ because both should behave in the same way given the same sequence of input.

\subsection{Cerberus Platform Invariants}

We describe a few key additional platform inductive invariants that were required to prove the \sre property on \tapc. Although the following list is not exhaustive\footnote{We refer the reader to the formal models in the \uclid code for the complete list of inductive invariants in full detail.}, it provides a summary of the difference between the invariants in the base \tap model and the \tapc model and explains what precisely makes the other sharing models more difficult to verify. The invariants are typically over the two traces $\pi_1$ and $\pi_2$ in the properties previously mentioned. However, there are single-trace properties, and unless otherwise noted, it is assumed that single-trace properties defined over a single trace $\pi$ hold for both traces $\pi_1$ and $\pi_2$ in the properties.

\paragraph{\textbf{Memory Sharing}}
We first begin with the invariants related to the memory-sharing model. As explained earlier, allowing the sharing of memory weakens the constraint that memory is strictly isolated. This means that the memory readable and executable by an enclave can either belong to itself or its root snapshot. This is true for the entrypoints of the enclave and the mapped virtual addresses. These are described as Eq.~(\ref{eq:ep_owned_by_self_or_rs}) and Eq.~(\ref{eq:memory_ownership}) respectively.

Eq.~(\ref{eq:ep_owned_by_self_or_rs}) states that all enclaves have an entrypoint that belongs to $e$ itself or its snapshot $\pi^i.\mdrs[e]$.
{\eqfontsize
    \begin{align}
    \label{eq:ep_owned_by_self_or_rs}
        \forall e\in\mathcal{E}_{id}, \forall i\in\mathbb{N}.\Big(& valid(e) \quad \Rightarrow\\\notag
        &\big(\pi^i.\towner[\pi^i.\mdep[e]] = e \quad \lor\\\notag
        &\pi^i.\towner[\pi^i.\mdep[e]] = \pi^i.\mdrs[e]\big) \Big)\\\notag
    \end{align}
}
\vspace{-0.7cm}

Eq.~(\ref{eq:memory_ownership}) states that every enclave $e$ whose physical address $p\in\PA$ corresponding to virtual address $v\in\VA$ in the page table that is mapped $\textit{mapped}_e(\pi[i].\tam[v])$\footnote{mapped is a function that returns whether a physical address is mapped in enclave $e$ and is equivalent to the \textit{valid} function in the CCS'17 paper~\cite{tap}} either belongs to $e$ itself or the root snapshot $\pi^i.\mdrs[e]$. 

To illustrate the potential complexity of the capped and arbitrary memory sharing models, the antecedent of this invariant would need to existentially quantify over all the possible snapshot enclaves that that own the memory as opposed to the current two (the enclave itself or its root snapshot). This would introduce an alternating quantifier\cite{reddy78} in the formula, making reasoning with SMT solvers difficult.

\vspace{-0.3cm}
{\eqfontsize
    \begin{align}
    \label{eq:memory_ownership}\notag
        \forall e\in & \mathcal{E}_{id}, v\in\VA, \forall i\in\mathbb{N}.\\\notag
        \Big(\big( & valid(e) \land active(e, \pi^i) \land mapped_e(\pi^i.\tam[v])\big) \Rightarrow \\
        \big(&\pi^i.\towner[\pi^i.\tam[v]] = e \quad \lor\\\notag
        &\pi^i.\towner[\pi^i.\tam[v]] = \pi^i.\mdrs[e]\big) \Big)
    \end{align}
}
Lastly, a memory that is marked free for an enclave $e$ is owned by that enclave itself, represented by Eq.~(\ref{eq:free_mem_owned}).
{\eqfontsize
    \begin{align}
    \label{eq:free_mem_owned}
        \forall e\in\mathcal{E}_{id}, p\in\PA, \forall i\in\mathbb{N}.\Big(\pi^i.\mdpaf[e][p] \Rightarrow \pi^i.\towner[p] = e \Big)
    \end{align}
}

\paragraph{\textbf{Snapshots}}
The next invariants relate to snapshot enclaves.
First, the root snapshot of an enclave is never itself as follows:
{\eqfontsize
    \begin{align}
    \label{eq:root_not_itself}
        \forall e\in\mathcal{E}_{id}, \forall i\in\mathbb{N}.(valid(e) \Rightarrow \pi^i.\mdrs[e] \neq e)
    \end{align}
}
Snapshots also do not have root snapshots Eq.~(\ref{eq:snapshot_no_rs}). This invariant reflects the property that the root snapshot to ancestor enclave relationship has a height of at most 1. This is stated as all enclaves that are snapshots have a root snapshot reference pointing to the invalid enclave ID $e_{inv}$.
{\eqfontsize
    \begin{align}
    \label{eq:snapshot_no_rs}
        \forall e & \in\mathcal{E}_{id}, \forall i\in\mathbb{N}.\\\notag
        \Big(&\big(valid(e) \land active(e, \pi^i) \land \pi^i.\mdiss[e]\big) \Rightarrow\\\notag
        &\pi^i.\mdrs[e] = e_{inv} \Big) &\\\notag
    \end{align}
}
\vspace{-0.7cm}

Next, if an enclave has a root snapshot that is not invalid \\(i.e. $\pi^i.\mdrs[e] \neq e_{inv}$), then the root snapshot is a snapshot and the child count is positive. This is represented as Eq.~(\ref{eq:root_not_itself_cc}).
{\eqfontsize
    \begin{align}
    \label{eq:root_not_itself_cc}
        \forall e\in\mathcal{E}_{id}, & \forall i\in\mathbb{N}.\\\notag
        \Big((valid & (\pi^i.\mdrs[e]) \land active(\pi^i.\mdrs[e], \pi^i)) \Rightarrow\\\notag
        \big(&\pi^i.\mdiss[\pi^i.\mdrs[e]] \quad\land\\\notag
        &\pi^i.\mdcc[\pi^i.\mdrs[e]] > 0\big) \Big)
    \end{align}
}
The last notable invariant says that the currently executing enclave cannot be a snapshot as described in Eq.~(\ref{eq:no_running_snapshots}).
{\eqfontsize
    \begin{align}
    \label{eq:no_running_snapshots}
        &\forall i\in\mathbb{N}.(\neg\pi^i.\mdiss[\pi^i.\tceid])
    \end{align}
}

We conclude this section by noting that coming up with the exhaustive list of inductive invariants for \tapc took a majority of the verification effort.

\section{Implementation in RISC-V Keystone}

To show its feasibility, we implement \cerberus on Keystone~\cite{keystone}.
Keystone is an open-source framework for building enclave platforms on RISC-V processors.
In Keystone, the platform operations $\mathcal{O}_{base}$ are implemented in high-privileged firmware called security monitor.
We implemented additional \snapshotop and \cloneop based on our specifications.
All fields of the enclave metadata are stored within the security monitor memory.
We extended the metadata with the variables corresponding to $\mdiss$, $\mdcc$, $\mdrs$, and $\mdpaf$.
All implementations are available at \url{https://github.com/cerberus-ccs22/TAPC.git}.

The implementation complies with the assumptions of the model described in \S\ref{sec:assumptions}.
First, Keystone enclave operations are atomic operations, and it updates the system state only when the operation succeeds.
Second, we implement a deterministic memory allocation algorithm for copy-on-write, by leveraging Keystone's free memory module.

For memory isolation, Keystone uses a RISC-V feature called Physical Memory Protection (PMP)~\cite{riscv-priv-110},
which allows the platform to allocate a contiguous chunk of physical memory to each of the enclaves.
When an enclave executes, the corresponding PMP region is activated by the security monitor.
We implemented the weakened constraints (i.e., Eq.~(\ref{eq:memory_ownership})) by activating the snapshot's memory region when the platform context switches into the enclave.

In the model, the platform would need to handle the copy-on-write.
In Keystone, an enclave can run with supervisor privilege, 
which allows the enclave to manage its own page table.
This was very useful when we prototype this work because the platform
does not need to understand the virtual memory mapping of the enclave.
Letting the enclave handle its own write faults does not hurt the security 
because the permissions on physical addresses are still enforced by the platform.
One implementation challenge was that the enclave handler itself would always trigger a write fault
because the handler requires some writable stack to start execution.
We were able to implement a stack-less page table traverse,
which allows the enclave to remap the page triggering the write fault without invoking any memory writes.
The final copy-on-write handler is similar to the on-demand fork~\cite{on-demand-fork}.

\section{Evaluation}

Our evaluation goals are to show the following:
\begin{itemize}
    \item \textbf{Verification Results:} Our incremental verification approach enables fast formal reasoning on enclave platform modifications.
    \item \textbf{Start-up Latency:} The \cerberus interface can be used with process-creation system calls to reduce the start-up latency of enclaves
    \item \textbf{Computation Overhead:} Our copy-on-write implementation does not incur significant computation overhead.
    \item \textbf{Programmability:} \cerberus provides a programmable interface, which can be easily used to improve the end-to-end latency of server enclave programs.
\end{itemize}

Throughout the performance evaluation, we used SiFive's FU540~\cite{HiFiveUnleashed} processor 
running at 1~GHz and 
an Azure DC1s\_v3 VM instance with an \Intel Xeon\textsuperscript{\textregistered}\xspace Platinum 8370C running at 2.4~GHz
to run Keystone and SGX workloads respectively.
Each experiment was averaged over 10 trials.

\subsection{Verification Results}
The \tapc model and proofs can be found at \url{https://github.com/cerberus-ccs22/TAPC.git}.

\par\textbf{Porting \tap from Boogie to \uclid}.
One other contribution of this work includes the port of the original \tap model from Boogie~\cite{boogie} to \uclid~\cite{uclid5memocode}. \uclid is a verification toolkit designed to model transition systems modularly, which provides an advantage over the previous implementation written in the software-focused verification IR Boogie. We also find that \uclid is advantageous over other state-of-the-art tools~\cite{dafny, z3smtsolver, cvc5, rosette} because of modularity and because it provides flexibility in modeling systems both operationally and axiomatically. This effort took three person-months working approximately 25 hours a week to finish.

\par\textbf{Verifying \tapc}.
The modeling and verification took roughly three person-months to write the extensions to the TAP model and verify using a scalable approach. We note that this time is substantially less than it would have taken to rebuild the model from scratch without an existing abstraction.

Fig.~\ref{fig:verify_result} shows the number of procedures \#pn, number of (uninterpreted) functions \#fn, number of annotations \#an (which include pre- and post-conditions, loop invariants, and system invariants), the number of lines of code \#ln. The last column shows the verification time which includes the time it took \uclid to generate verification conditions and print them out in SMTLIB2.0 and verify them using Z3/CVC4\footnote{For one of the properties, we found that Z3 would get stuck but CVC4 didn't.}. The time discrepancy between the original proofs ~\cite{tap} and the ones in this effort can be explained by the way we generate all the verification conditions as SMTLIB on disk before verifying as a way to use other SMT solvers. We also use \uclid instead of Boogie~\cite{boogie}. We note that the number of lines for \snapshotop and \cloneop is 1110, which means only 489 lines were used to extend the existing \tap operations and platform model.

Despite the added complexity, each operation for each proof took only a few minutes to verify individually as shown in the last column of Fig.~\ref{fig:verify_result}. This demonstrates that our incremental verification methodology is practical and consequently reduces the overall time to verify additional operations at a high level.

\begin{figure}[t]
\centering
    \footnotesize
    \begin{tabular}{ p{0.1\textwidth} p{0.03\textwidth} p{0.03\textwidth} p{0.03\textwidth} p{0.03\textwidth} c }
    \specialrule{.1em}{.1em}{.1em} 
     \multirow{2}{*}{\textbf{Model/Proof}} & \multicolumn{4}{c}{\centering\textbf{Size}} & \multirow{2}{*}{\vtop{\hbox{\strut \textbf{Verif.}}\hbox{\strut \textbf{Time (s)}}}} \\
     & \#pr & \#fn & \#an & \#ln\\
     \hline
     \multicolumn{6}{c}{\textbf{\tap Models}}\\
     \hline
     \texttt{TAP} & 43 & 14 & 225 & 2100 & 140\\
     \texttt{Integrity} & 2 & 0 & 52 & 525 & 285\\
     \texttt{Mem. Conf} & 3 & 0 & 44 & 838 & 342\\
     \hline
     \multicolumn{6}{c}{\textbf{\tapc Models}}\\
     \hline
     \texttt{TAP} & 45 & 16 & 466 & 3689 & 1380\\
     \texttt{Integrity} & 2 & 0 & 109 & 937 & 934\\
     \texttt{Mem. Conf} & 3 & 0 & 119 & 1307 & 944\\
    \specialrule{.1em}{.1em}{.1em}
    \end{tabular}
    \caption{Model Statistics and Verification Times}
    \label{fig:verify_result}
\end{figure}

As evident by the results, we emphasize that the single-sharing model is practical to formally encode and verify. We also confirm that introducing invariants with alternating quantifiers and existential quantifiers in our models degraded the verification time and would likely do the same for alternative models. These attempts heavily influenced our decisions and we strongly encourage the use of our sharing model.

\subsection{Start-up Latency}

\begin{figure}
\lstset{language=C,
                basicstyle=\ttfamily\footnotesize,
                keywordstyle=\color{blue}\ttfamily,
                stringstyle=\color{red}\ttfamily,
                commentstyle=\color{green}\ttfamily,
                morecomment=[l][\color{magenta}]{\#}
}
\begin{lstlisting}[frame=single]
int main() {
  char* buf = malloc(SIZE);
  clock_t start = clock();
  if (!fork()) {clock_t end = clock();} // child
  else { return; }                      // parent
}
\end{lstlisting}
\caption{C code to measure \texttt{fork} latency}
\label{fig:fork_latency_code}
\end{figure}

To show the efficacy of \cerberus interface, we implement \sysfork and \sysclone system calls based on \cerberus.
When the system calls are invoked in the enclave program, it calls \snapshotop to create an immutable image and cooperates with the OS to clone two enclaves from the snapshot using \cloneop.
We compare the latency of \sysfork on two different platforms: SGX-based Graphene~\cite{graphene-sgx} (now Gramine Linux Foundation project~\cite{gramine}) and RISC-V Keystone~\cite{keystone} with \cerberus. 
Fig.~\ref{fig:fork_latency_code} shows the program that calls \sysfork after allocating memory with \texttt{SIZE}.

\begin{figure}
\centering
	\includegraphics[width=0.7\columnwidth]{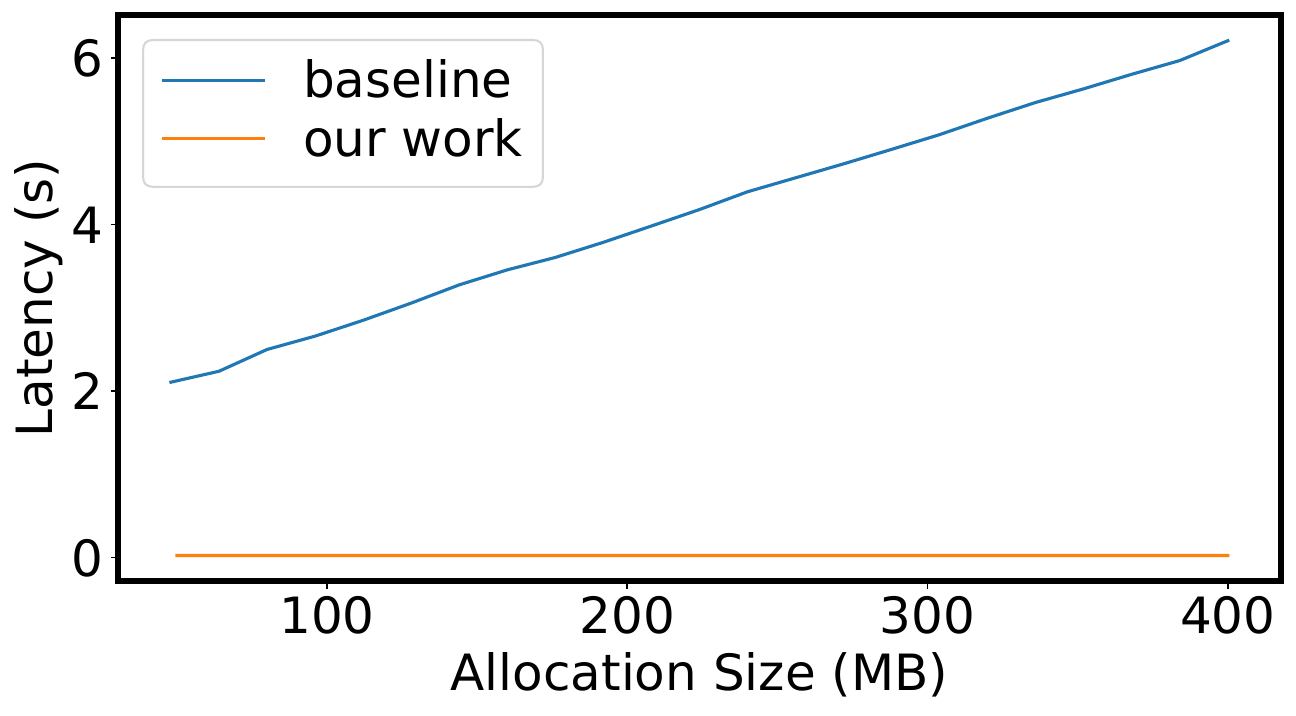}
	\caption{The latency of \sysfork with respect to the size of the allocated memory.}
	\label{fig:eval_exp2}
\end{figure}

The baseline (Graphene-SGX) latency increases significantly as the allocation size increases (Figure~\ref{fig:eval_exp2}).
With a 400~MB buffer, it takes more than 6 seconds to complete.
Also, each of the enclaves will take 400~MB of memory at all times, even when most of the content is identical until one of the enclaves writes.
With \cerberus, the latency does not increase with respect to the allocation size.
This is because we are not copying any of the parent's memory including the page table.
It only took 23 milliseconds to fork on average, with a standard deviation of 16 microseconds.

\subsection{Computation Overhead}

\begin{figure}
	\includegraphics[width=1\columnwidth]{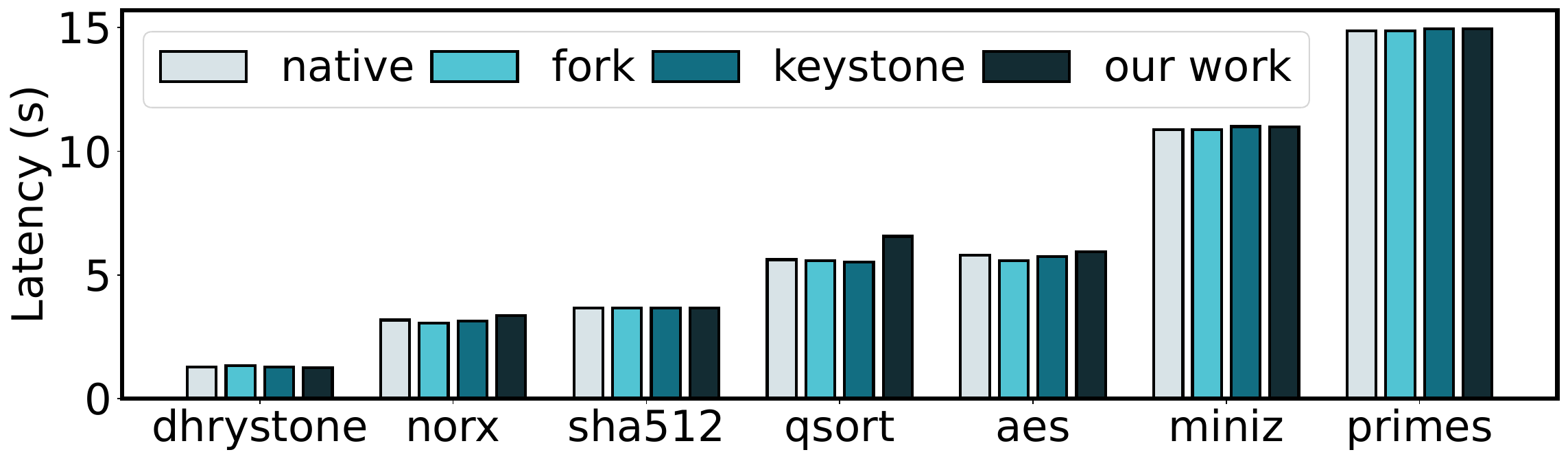}
	\caption{Computation Overhead on RV8. \textbf{native}: native execution of the original RV8, \textbf{fork}:
	native execution of the modified RV8 with \sysfork, \textbf{keystone}: enclave execution of the original RV8,
	and \textbf{our work}: enclave execution of the modified RV8 with \cerberus.}
	\label{fig:cow-overhead}
\end{figure}

We measure the computation overhead incurred by CoW invocation.
In order to see the overhead for various memory sizes and access patterns, we use RV8~\cite{rv8} benchmark.
RV8 consists of 8 simple applications that perform single-threaded computation.
We omit \texttt{bitint} as we were not able to run it on the latest Keystone, because of a known bug on their side.
Since RV8 does not use the \sysfork system call, we have modified RV8 such that each of them forks before the computation begins.
Note that all of the application starts with allocating a large buffer, so we inserted \sysfork after the allocation.
Thus, copy-on-write memory accesses are triggered during the computation depending on the memory usage.

As you can see in Figure~\ref{fig:cow-overhead}, the average computation overhead of copy-on-write memory over Keystone was only 3.9\%.
The worst overhead was 19.0\% incurred in \texttt{qsort}, which uses the largest memory (about 190 MiBs).
We argue that the benefit of cloning an enclave is small
for such workloads that have a large buffer that is not shared across enclaves.

\subsection{Programmability}

To show the programmability of \cerberus interface, we showcase how server programs can leverage memory sharing to improve their end-to-end performance.

Although \snapshotop or \cloneop are not directly related to \sysfork or \sysclone, their behavior maps well with \snapshotop and \cloneop.
For example, those system calls create a new process with exactly the same virtual memory, which can be mapped to \cloneop and optimized by \snapshotop.
Thus, we provided two co-authors with the modified \sysfork and \sysclone that use \cerberus interface and asked to make the server programs leverage memory sharing.

An author modified darkhttpd, a single-threaded web server, to fork processes to handle new HTTP requests inside the event loop. 
This allowed darkhttpd to serve multiple requests concurrently and continue listening for new requests.
We measure the latency of an HTTP request using \texttt{wget} to fetch 0.5 MB of data.
The resulting program incurs only a 2.1x slowdown over the native (non-enclave) execution, in contrast to a 33x slowdown in corresponding Intel SGX implementation (the exact same program ran with Graphene).
The 2.1x overhead is mainly due to the slow I/O system calls, which is a well-known limitation of enclaves~\cite{weisse2017hotcalls, keystone}.

Another author implements a simple read-only database server application using \textit{Sqlite3}, which is a single-file SQL library that supports both in-memory and file databases.
The resulting program serves each query with a fresh child created by \sysfork.
We measure the latency of 1,000 \texttt{SELECT} queries served by separate enclaves.
The resulting program incurs a 36x slowdown over the native execution, compared to a 262x slowdown in corresponding Intel SGX implementation.
SGX overhead is much worse than in Darkhttpd because there is more data to copy over (the entire in-memory database).
The 36x slowdown is mainly due to limited concurrency in Keystone: since Keystone implements memory isolation with a limited number of PMP entries, it can support only up to 3-4 concurrent enclaves.
This is not an inherent limitation of \cerberus.

Both authors did not have any difficulties in allowing enclaves to share a memory, because they were already familiar with the expected behavior of the system calls.
However, they did not have any knowledge of the codebase of Darkhttpd nor Sqlite3 prior to the modification.
Darkhttpd required modification of less than 30 out of 2,900 lines of code, which took less than 10 person-hours, and Sqlite3 consists of 103 lines of code, which took less than 20 person-hours.
This shows that the \cerberus extension can be easily used to improve the end-to-end performance of server programs.
\section{Discussion}
\label{sec:discussion}

\paragraph{Low-Equivalent States}
Our security model contains the notion of low states as in standard observational determinism type properties, even though they are not explicitly stated in section \S~\ref{sec:guarantees}. Instead, the low states are constrained to be equal in the antecedent of the implication of the properties.
The traditional non-interference or observational determinism properties most closely resemble the confidentiality Eq.~\ref{eq:confidentiality}.
In this property, the low states include the inputs $I_{e_1}(\sigma), I_{e_2}(\sigma)$ to the enclaves and platform operations controlled by the untrusted OS, all of the adversary
controlled enclaves' state  $E_e(\sigma)$, where $e$ is not the protected enclave (i.e., $e_1, e_2$), and the adversary state $A_{e_1}(\sigma), A_{e_2}(\sigma)$. The idea
is to prove that under the same sequence of adversarial controlled inputs, the adversary cannot differentiate between the two traces which have the same enclave (i.e. $e_1$, $e_2$) with differing high data in memory.
Similarly, the low states of the integrity property (Eq.~\ref{eq:integrity}) includes the untrusted inputs $I_{e_1}(\sigma), I_{e_2}(\sigma)$. However, instead of constraining the adversary inputs
to be the same, we want to show that an enclave executes deterministically regardless of the state outside the enclave. As a result, the remaining low states include the protected enclaves
$E_{e_1}(\sigma), E_{e_2}(\sigma)$.
Lastly, secure measurement can be viewed as a form of integrity proof and contains the same low states as the integrity property.

\paragraph{Performance Comparison with Previous Work}
The evaluation section does not make direct performance comparisons with previous work such as PIE, which is based on x86.
However, based on our calculations, \cerberus's overhead is on par with PIE. 
For example, PIE incurs about 200ms startup latency on serverless workloads~\cite{pie} whereas Cerberus incurs 23ms on \sysclone system calls. 
We analyze that \cerberus is faster mainly because it leverages Keystone’s ability to quickly create an enclave with zero-filled memory without measurement, which SGX does not support. 
PIE’s copy-on-write introduces 0.7-32ms overhead on serverless function invocations taking 144-1153ms (the paper did not provide relative overhead over native execution), which can be roughly translated into less than a few percent of overhead, which is similar to \cerberus.

\paragraph{Verifying the Implementation}
This paper does not verify the implementation of \cerberus in Keystone.
Unsurprisingly, any discrepancy between the model and the implementation can make the implementation vulnerable.
In particular, the enclave page table is abstracted as enclave metadata in \tap and \tapc, where it is actually a part of memory $\tmem$.
\cerberus in Keystone does not create any security holes because the page table management is trusted (the enclave manages it). 
However, this does not mean that we can apply the same argument to the other implementations.
To formally verify the implementation, we can construct the model for Keystone implementation
and do the refinement proof to show that the model refines the TAP model
as described by Subramanyan \etal~\cite{tap}.
We leave this as future work.

\paragraph{In-Enclave Isolation}
\label{sec:in_enclave_isolation}

Instead of modifying the platform, 
a few approaches~\cite{occlum,chancel,enclavedom,sgxbounds} use
\textit{in-enclave} isolation to
create multiple security domains within a single enclave.
However, security guarantees of such solutions rely on the formal properties of not only the enclave platform,
but also the additional techniques used for the isolation.
For example, the security of software fault isolation (SFI)~\cite{sfi} based approaches~\cite{occlum,chancel} depends on the correctness and robustness of the SFI techniques
including the shared software implementation and the compiler, which should be formally reasoned together with the enclave platform.
Thus, such approaches will result in a significant amount of verification efforts.

\section{Conclusion}

We showed how to formally reason about modifying the enclave platform to allow memory sharing.
We introduce the single-sharing model, which can support secure and efficient memory sharing of enclaves.
We also proposed two additional platform operations similar to existing process-creation system calls.
In order to formally reason about the security properties of the modification, 
we defined a generic formal specification by incrementally extending an existing formal model.
We showed that our incremental verification allowed us to quickly prove the security guarantees of the enclave platform.
We also implemented our idea on Keystone open-source enclave platform and 
demonstrated that our approach can bring significant performance improvement to server enclaves.

\section{Acknowledgments}

We thank the anonymous reviewers for their insightful comments.
This work was supported in part by the Qualcomm Innovation Fellowship, by Amazon, by Intel under the Scalable Assurance program, and by RISE, ADEPT, and SLICE Lab industrial sponsors and affiliates.
Any opinions, findings, conclusions, or recommendations in this paper are solely those of the
authors and do not necessarily reflect the position or the policy of the sponsors.

\bibliographystyle{ACM-Reference-Format}
\balance
\bibliography{main}

\end{document}